\documentclass{article}

\usepackage{chicago}
\usepackage{amsmath, amsfonts, amsthm}
\newcommand{\ie}{\mbox{i.\,e.\,\ }}
\newcommand{\iec}{\mbox{i.\,e.\,}}

\newcommand{\egc}{\mbox{e.\,g.\,}}

\newcommand{\be}{\begin{equation}}
\newcommand{\ee}{\end{equation}}

\newcommand{\dr}[1]{\ensuremath{\mathrm{d} #1\,}}
\newcommand{\mc}[1]{\ensuremath{\mathcal{#1}}}

\newcommand{\vctr}[1]{\ensuremath{\mathbf{ #1 }}}

\newcommand{\bea}{\begin{eqnarray}}
\newcommand{\eea}{\end{eqnarray}}
\newcommand{\bi}{\begin{itemize}}
\newcommand{\ei}{\end{itemize}}
\newcommand{\ben}{\begin{enumerate}}
\newcommand{\een}{\end{enumerate}}
\newcommand{\bq}{\begin{quote}}
\newcommand{\eq}{\end{quote}}
\newcommand{\bd}{\begin{description}}
\newcommand{\ed}{\end{description}}
\newcommand{\bdm}{\begin{displaymath}}
\newcommand{\edm}{\end{displaymath}}
\newcommand{\ba}{\begin{abstract}}
\newcommand{\ea}{\end{abstract}}

\renewcommand{\d}{\partial}
\newcommand{\field}[1]{\mathbb{#1}}

\renewcommand{\ss}{\ensuremath{W}}
\begin{document}

\title{Empirical consequences of symmetries}
\author{Hilary Greaves\thanks{Somerville College, Oxford, and Philosophy Faculty, University of Oxford} \\ David Wallace\thanks{Balliol College, Oxford, and Philosophy Faculty, University of Oxford}}
\date{March 28, 2011}
\maketitle

\begin{abstract}
`Global' symmetries, such as the boost invariance of classical mechanics and special relativity, can give rise to direct empirical counterparts such as the Galileo-ship phenomenon. However, a widely accepted line of thought holds that `local' symmetries, such as the diffeomorphism invariance of general relativity and the gauge invariance of classical electromagnetism, have no such direct empirical counterparts. We argue against this line of thought. We develop a framework for analysing the relationship between Galileo-ship empirical phenomena and physical theories that model such phenomena that renders the relationship between theoretical and empirical symmetries transparent, and from which it follows that both global and local symmetries can give rise to Galileo-ship phenomena. In particular, we use this framework to exhibit analogs of Galileo's ship for both the diffeomorphism invariance of general relativity and the gauge invariance of electromagnetism.
\end{abstract}

\section{Introduction}
\label{intro}
There is something of a tension between two aspects of symmetry in physics. On the one hand, there is a widespread consensus\footnote{See, \egc, \citeN{saundersleibniz} and references therein.} that `two states of affairs related by a symmetry transformation are really just the same state of affairs differently described': that is, that if two mathematical models of a physical theory are related by a symmetry transformation, then those models represent one and the same physical state of affairs. This seems to give symmetry a purely formal role in physics: symmetries are not features of the world, but merely features of our method of describing the world.

On the other hand, it seems to be a matter of plain historical fact that the observable consequences of symmetries have guided physicists in their construction of theories ever since Galileo's ship thought experiment. It is \emph{prima facie} mysterious how this fact is compatible with the account of symmetries given above.

The mystery is solved, or at least demoted from mystery to puzzle, once we recognise that a transformation can be a symmetry of a \emph{subsystem} of the world without being a symmetry of the \emph{whole} world.\footnote{This observation has been made by (amongst others) \citeN{brownsypel}, \citeN{Kosso2000}, \citeN{bradingbrownsymmetry} and \citeN{Healey2009}.}
As such, the transformation gives rise to a different state of affairs, but has no detectable consequences for measurements confined to the subsystem.

For example, Galileo famously noted that, provided the interior of a ship is suitably dynamically isolated from goings-on outside, experiments inside the cabin of a ship could not tell whether the ship was stationary or moving with uniform velocity with respect to the shore:
\bq
Shut yourself up with some friend in the main cabin below decks on some large ship \ldots With the ship standing still, observe carefully [the results of various experiments you conduct inside the cabin. Then] have the ship proceed with any speed you like,
so long as the motion is uniform and not fluctuating this way and that. You will discover
not the least change in all the effects named, nor could you tell from any of them whether
the ship was moving or standing still. \cite[pp.\,186--7]{galileotwoworldsystems}
\eq
Here, the subsystem is the ship's cabin, and the transformation is a uniform boost of the ship relative to the shore. Restricted to the cabin, the transformation is a symmetry, and as such, has no detectable consequences as long as the cabin remains dynamically isolated from the exterior. But it is not a symmetry of the world as a whole: measurements in which the cabin interacts with the outside world (such as those made by opening the window and looking out) will in general have results which depend on whether or not the transformation has been carried out.

This line of thought \emph{seems} to suggest an important distinction between ``global'' symmetry transformations, in which the same transformation must be carried out at each point of space and time in order for the transformation to be a symmetry  (like translations and boosts in special relativity or global phase transformations in Klein-Gordon theory) and ``local'' symmetry transformations, in which the transformation may vary arbitrarily from point to point (as in  the gauge symmetry of electromagnetism or the diffeomorphism symmetry of general relativity). \emph{Prima facie}, any transformation which is a local symmetry of each subsystem of a system is also a local symmetry of the system as a whole, and hence does not relate different states of affairs to one another: hence, it seems (\emph{prima facie}), we should expect that whilst global symmetries have direct empirical significance, local symmetries do not.

\citeN{bradingbrownsymmetry}, \citeN{healeybook}, and \citeN{Kosso2000} all defend this view that local symmetries cannot have direct empirical significance. Brading and Brown, in fact, go further and argue that only global \emph{spacetime} symmetries (as opposed to global internal symmetries like global phase transformations)  can have direct empirical significance.

There are at least three reasons to be deeply puzzled by these claims, though. First: in any theory that has a `local' symmetry group, the `global' symmetries remain as a subgroup of that local symmetry group. (For example, in general relativity, the `global' translations and boosts form a subgroup of the group of all diffeomorphisms.) It is therefore logically impossible that all global symmetries, but no local symmetries, can have direct empirical significance; and it would be highly mysterious if global symmetries managed to have empirical significance \emph{while no other symmetries were around}, but somehow lost this capacity once the full local group of transformations appeared as symmetries.

Secondly (and relatedly): very often in the history of physics, a theory with a `global' symmetry is superseded by a more accurate theory in which that symmetry is `localised': thus, for example, the (global) Poincar\'{e} symmetry group of a special-relativistic theory is extended to the (local) diffeomorphism group of general relativity, and the (global) group of constant shifts in electric potential in electrostatics is extended to the (local) group of `gauge' transformations of the four-potential in electrodynamics. It would be extremely odd (at best) if we had a theoretical explanation of an empirically observable symmetry while we stuck with the less accurate theory, but nothing of this explanation remained once we passed to the larger symmetry group in the more accurate theory: if for example, there was nothing we could say \emph{from the perspective of general relativity} to explain Galileo's ship in terms of boost invariance. Scientific realists should regard this as not only odd, but also rather worrying: if the explanatory successes of old theories aren't reproduced in their successors, much of the justification for realism is undermined.

Thirdly: if the relationship between the boost invariance of Newtonian mechanics and the Galileo-ship thought experiment provides the defining example of the sense in which some theoretical symmetries may have `direct empirical significance', we can (and will) sketch several examples that \emph{seem}, \emph{prima facie} at least, to be perfect analogs of the Galileo-ship scenario for cases of local symmetry.

Our purpose in this paper is to present and illustrate a new framework for thinking about  the empirical consequences of   symmetries which  clears up these confusions.  We will show that although a distinction related to the global/local distinction does determine whether or not a symmetry has direct empirical significance \emph{in a particular empirical situation}, nothing like the global/local distinction tells us which symmetries can \emph{in general} have empirical significance --- and in particular, it is false that local symmetries are in general unobservable.  In particular (we will argue),  there are analogs of the Galileo-ship scenario for both diffeomorphism and gauge symmetries.

The structure of the paper is as follows. Section \ref{faraday_thooft} introduces two examples --- that of Faraday's Cage, and a modification of the two-slit experiment --- that appear to be just such analogs of the Galileo-ship scenario for cases of local symmetry, and \emph{registers} our disagreement with previous discussions (respectively, \citeN{Healey2009} and \citeN{bradingbrownsymmetry}) that claim to find relevant disanalogies, and hence to support the conventional wisdom in the face of such examples. To \emph{justify} that disagreement requires us to develop a comprehensive theory and classification of the various circumstances under which one can perform a symmetry transformation on a subsystem without thereby performing a symmetry transformation on
the universe, and to work out its application to a variety of examples; this task occupies the larger part of the paper. Section \ref{framework1} begins to set up our framework, including the ways in which we regiment talk of symmetries, subsystems, environments, and boundary conditions governing compatibility between given subsystem and environment states. Section \ref{coulombic_electrostatics} applies this initial part of the abstract framework to the symmetries and subsystems of electrostatics, by way of illustrative example. Section \ref{framework2} lays out the remainder of our framework, and defines the notions of `interior' and `boundary-preserving' symmetries, which we believe are crucial to an understanding of the phenomenon of Galileo-ship type scenarios. The next four sections apply the completed framework to four other theories. Section \ref{newtonian_gravity} discusses Newtonian gravity, and identifies a particular subgroup of that theory's symmetry group as corresponding to the `Einstein's elevator' empirical symmetry. Up to this point, our examples will have included only `global' symmetries; sections \ref{electromagnetism}--\ref{gr} deal with `local' symmetries. Section \ref{electromagnetism} discusses classical electromagnetism, and identifies the empirical symmetry of Faraday's cage as an empirical consequence of the local gauge symmetry of that theory. Section \ref{kgm} discusses Klein-Gordon-Maxwell theory, and identifies the empirical symmetry in the modified two-slit setup that we first discuss in section \ref{faraday_thooft} as an empirical consequence of the local gauge symmetry of \emph{that} theory.  Section \ref{gr} discusses general relativity, and identifies both Galileo's ship and Einstein's elevator as empirical consequences of particular subgroups of the diffeomorphism symmetry group of GR. Section \ref{summary} contains a compact presentation of our finished abstract framework, and a summary of our claims about the conditions under which symmetries can and cannot have empirical significance. Section \ref{conclusion} is the conclusion.

\section{Analogs of Galileo's ship? Faraday's cage and t'Hooft's beam-splitter}
\label{faraday_thooft}

 \citeN{Healey2009} presents a puzzle for the `conventional wisdom' about local and global symmetries; his puzzle will serve as a starting point for our discussion. To explain the puzzle --- and to begin theorising about the empirical consequences of symmetries --- we follow Healey in distinguishing   between \emph{empirical symmetries} and \emph{theoretical symmetries}.

 Theoretical symmetries are automorphisms of the class of mathematical models of a theory. For our purposes, we can \emph{define} a theoretical symmetry to be a map $f:\Theta \rightarrow \Theta$ from the set $\Theta$ of models of a theory $T$ to itself, such that any two models $m,f(m)$ related by that map represent the same  physical state of affairs --- the same possible world, if preferred  --- as one another. An example of a group of theoretical symmetries is given by the boost invariance of Newtonian mechanics: we work with a formulation of Newtonian mechanics that uses coordinate systems, and a transformation $\mathbf{x} \mapsto \mathbf{x} - \mathbf{v}t$ for the whole universe results merely in a redescription of the same physical state of affairs in a different inertial frame (that is, there is no distinction between `active' and `passive' transformations when the transformation is performed on the universe as a whole).\footnote{This view commands widespread if perhaps not universal agreement in physics and philosophy of physics; in this paper we assume it without further discussion. (For a positive case, see \citeN{saundersleibniz}.) Those unsympathetic to the view, however, can replace `represent the same possible world' in our definitiion of `theoretical symmetry' with `represent an experimentally indistinguishable possible world'; \emph{mutatis mutandis}, the rest of our discussion will go through unimpeded.}

An \emph{empirical} symmetry is an altogether different sort of thing: one has an empirical symmetry only if one has two \emph{physically} distinct states of the world, which (however) cannot be discriminated from one another by means of measurements confined to some specified subsystem. \footnote{Healey proceeds at this point by introducing talk of sets of `situations' (to be thought of as possible physical states of affairs in some proper part of the universe, and sharply contrasted with mathematical models of theories), and identifying empirical symmetries with automorphisms of such sets of situations. We won't need this development.}
An example of an empirical symmetry is the boost of Galileo's (physical) ship by a given uniform velocity $\mathbf{v}$. This empirical symmetry `corresponds to' the theoretical boost invariance of Newtonian mechanics in the sense that it follows from the assumptions (i) that Newtonian mechanics adequately describes the Galileo ship scenario, (ii) that Newtonian mechanics is boost invariant and (iii) that the ship's cabin is effectively dynamically isolated\footnote{``Dynamical isolation'' is of course approximate in practice, since no proper subsystem can in practice be \emph{perfectly isolated}. It is also relative to the observation capabilities of relevant observers (if Galileo had included a GPS tracker in his list of the cabin's accoutrements, things would have been rather different).} from the environment relative to which it is boosted, that experiments confined to the cabin cannot distinguish between the stationary and moving states of the ship.

\subsection{Faraday's cage}
\label{healeys_puzzle}

Healey supports the conventional wisdom that `local' symmetries cannot have direct empirical significance in this sense: that `there can be no analogue of the Galilean ship experiment for local gauge transformations' \cite[p.657]{bradingbrownsymmetry}. If this is the case, his puzzle is that the following case of \emph{Faraday's cage} seems, \emph{prima facie} at least, to \emph{be} a perfect analog in all relevant respects.

Faraday's observation was that, if one constructs a cage from conducting material and places a static charge on the walls of the cage, then experiments confined to the interior of the cage cannot detect the presence or magnitude of that charge. Physically, this is because, in the limit in which the material is \emph{perfectly} conducting, charge placed on its exterior will redistribute itself under the influence of electrostatic repulsion, and the equilibrium distribution is such that the net electric field at any point inside the cage --- summing contributions from from the charges everywhere on the surrounding wall --- is zero.

Constructing such a cage and performing the experiment, Faraday writes

\bq
I went into this [charged] cube and lived in it, but though I used lighted candles, electrometers, and all other tests of electrical states, I could not find the least influence upon them. \cite[p.53]{Maxwell1881}
\eq

The puzzle arises because, in electromagnetic theory, charging the cage has the effect of \emph{raising its  electric potential} --- but shifts of  electric potential are elements of a \emph{local} symmetry group of electromagnetism.
Hence Healey's puzzle: It \emph{seems} that the empirical symmetry between Faraday's charged and uncharged cages is a consequence of the theoretical gauge symmetry of electromagnetism, in precisely the same way as that in which the empirical symmetry between Galileo's stationary and moving ships is a consequence of the boost invariance of Newtonian mechanics.
The challenge to an advocate of the conventional wisdom is therefore: `why doesn't Faraday's cube provide a \emph{perfect} analogue
of Galileo's ship for a class of local gauge transformations of classical electromagnetism?'\cite[p.701; emphasis in original]{Healey2009}.

Healey's own view  is that, the striking \emph{prima facie} analogy between Faraday's cages and Galileo's ships notwithstanding, there is in fact an important disanalogy between the two, and (furthermore) it is a disanalogy that upholds the conventional wisdom: he concludes from it that
`[l]ocal gauge symmetry is \emph{merely} a theoretical symmetry
of classical electromagnetism; and [implies] no corresponding empirical gauge symmetry' \cite[p.710; emphasis added]{Healey2009}. We disagree: there \emph{is} a disanalogy between Galileo's ship and Faraday's cage that is of some relevance to the empirical status of the theoretical symmetries involved, but it is not the disanalogy that Healey picks out. Further, we will argue,  the relevant disanalogy that there \emph{is}  has nothing to do with the global/local distinction, and does not imply that the gauge symmetry of electromagnetism has less (or less direct) empirical significance than the boost invariance of Newtonian mechanics.\footnote{The present paper is devoted to laying out our preferred framework for thinking about the empirical significance of symmetries, and drawing out its consequences for various cases. If our analysis is correct, it follows that Healey's must be incorrect, but in this paper we do not attempt to say where.}

\subsection{t'Hooft's beam-splitter}
\label{thooft}

A similar puzzle is considered by \citeN{bradingbrownsymmetry}, in discussion of \citeN{thooft}. They
consider the following variant on the two-slit experiment. A beam (say, of electrons) is split in two. The two halves of the beam follow disjoint paths until they are recombined some time later. The resulting interference pattern is observed by having the beam stopped by a fluorescent screen. On each run of the experiment, the experimenter may or may not insert a 180 degree phase-shifter into the upper, but not the lower, path. It is observed that the interference pattern depends nontrivially on whether or not the phase-shifter is present.

Our interest in this setup\footnote{This was not t'Hooft's interest. t'Hooft introduced the setup as a purported empirical illustration of the fact that `gauge transformations' on the wavefunction alone---transformations of the form
\bea
\psi^\prime(x) & = & e^{i \theta(x)} \psi(x) \\
A^\prime_\mu(x)  & = & A_\mu(x)
\eea
---are not symmetries of the theory. That it is an empirical illustration of that fact is contested by Brading and Brown.} arises from the idea that one can treat the upper half of the beam as subsystem (analogous to Galileo's ship), and the lower half as environment (analogous to Galileo's shore). We know that by performing experiments on the upper half of the beam (subsystem) \emph{alone}, we would not be able to detect the presence or absence of the phase-shifter; it is only when we allow the upper half to interact with the lower half (environment) that we can detect whether or not the phase-shift has been performed.

But this setup can be modelled in a theory that represents the state of the electron via a complex scalar field, $\psi$, coupled to an electromagnetic field that is represented by a four-potential $A_\mu$ (see section \ref{kgm}).\footnote{The complex scalar field is generally taken to be a one-particle wavefunction, and may be so taken in our discussion. However, the empirical results discussed would also occur in a possible world containing a classical charged field, in which our analysis would also apply; one of us has argued \cite{wallaceantimatter} that for this reason there is nothing essentially quantum about the AB effect.} The following `gauge transformations' are symmetries of the theory:
\be
\begin{array}{rcl}
\psi & \mapsto & e^{i \chi} \psi, \\
A_\mu & \mapsto & A_\mu + \d_\mu \chi,
\end{array}
\ee
where, again, $\chi$ is an arbitrary smooth function on spacetime (but now the transformation must modify \emph{both} $\psi$ and $A_\mu$ in the manner stated, with common $\chi$, in order to be a symmetry).

But the phase-shifter precisely implements a transformation $\psi \mapsto e^{i \theta} \psi$ on the upper half-beam while leaving the electromagnetic potential unchanged (as gauge transformations with constant $\chi$ do). Hence Healey's question arises again for this case: Why doesn't the modified two-slit experiment provide a \emph{perfect} analog of the Galileo-ship thought experiment, vis-a-vis the notion of direct empirical significance, for local gauge symmetry?

Brading and Brown agree with Healey: the analogy is superficial, and the empirical result is not --- despite appearences --- a direct consequence of the theoretical symmetry. In this case as in the last, we disagree: indeed, our view is that t'Hooft's beam-splitter, even more so than Faraday's cage, is a \emph{direct} analog of Galileo's ship. To defend this view  requires us to lay out our preferred framework for thinking about the  empirical consequences of symmetries; it is to this task that we now turn.

\section{A framework for symmetries I --- systems and subsystems}
\label{framework1}

If the observable consequences of a theory's symmetry transformations arise from performing those transformations on subsystems of a system, one might think that a careful analysis of this process ought to start by giving explicit definitions of `subsystem' and `environment'. The language of differential geometry, in particular, seems especially well suited to this (``given a model $\langle \mc{M},g,\phi\rangle$ of the theory, a subsystem of that model is an open subset of $\mc{M}$ together with the restrictions of $g$ and $\phi$ to that subset, such that \ldots''). But this sort of explicit move falls foul of the open-ended variety of subsystems that physics provides us with. The stream of neutrinos passing through your body at this moment, for instance, is a system interacting only very weakly with you despite its spatial coincidence: a fact which allows us (in principle!) to demonstrate the Lorentz symmetry group empirically by performing a Lorentz boost on you but not the neutrinos. A single electron can with profit be treated as comprising two systems --- specified by the spin and spatial degrees of freedom --- which in many circumstances can be treated as effectively isolated. And in unitary quantum mechanics, `system' and `environment' become branch-relative concepts in a sense that is not easily characterised in terms of spacetime regions.

For this reason, we choose instead to adopt a more axiomatic approach: we define a framework into which all the examples above (and, hopefully, any others) can fit, and we take this framework \emph{implicitly} to define `subsystem' and `environment' for present purposes. We suppose firstly that there exists a set \mc{S} of states of the subsystem, and another set \mc{E} of states of the environment. These `states' are intended to be mathematical objects: the mathematical entities representing the physical state of system and environment, not those physical states themselves. In a field-theoretic context, for instance, states of \mc{S} might be specified by the field values on some subset $\mc{N}$ of the theory's underlying manifold \mc{M}, and states of \mc{E} would be specified by the field values on the rest of the manifold.

It would be natural to define states of the total system as ordered pairs $\langle s,e\rangle$ of states $s\in \mc{S}$ and $e\in \mc{E}$, and we shall do just this. But there are three important caveats.

Firstly, in doing so we make the assumption that knowing the state of the subsystem and its environment suffices to specify the state of the total system. There is a \emph{prima facie} worry that this assumption will exclude such phenomena as quantum entanglement between `subsystem' and `environment'.  However, this worry  is misguided: our assumption is one about the mathematical formalism with which we describe the theory and not about the theory itself. For example, it is true for Yang-Mills gauge theories in the connection formalism but not in the holonomy formalism; it is true for quantum mechanics in (Deutsch and Hayden's construal of) the Heisenberg picture but not in the Schr\"{o}dinger picture.\footnote{For more information on the different formalisms for Yang-Mills theories, see \citeN{healeybook} and references therein; for discussions of locality in different formulations of quantum  theory, see \citeN{deutschhayden} and \citeN{wallacetimpsonshort}.} 

Secondly, the idea that subsystem and environment can be treated as distinct and isolated --- so that $\langle s,e\rangle$ and $\langle s',e\rangle$,  can both be treated as possible states of the total system --- is at best an idealisation. Forces in the physical universe are not normally so cooperative as \emph{entirely} to screen off any system from another; the gravitational field of an arbitrary planet in Andromeda has a non-zero  dynamical effect on your body.  But in a great many situations of physical significance, the idealisation is highly accurate.

Finally, and of most direct importance for this paper, note that one should not assume that just any old pair $\langle s,e\rangle$ is a possible state of the total system. In general, there are boundary conditions which must be satisfied in order for a given subsystem-environment pair to be a solution of the theory's dynamical equations. In the case where the total system is a field theory, these will be boundary conditions in the differential-equation sense: requirements that the fields and their derivatives match up on the boundary. In other cases they may take a different form: in Heisenberg-picture quantum mechanics, for instance, the boundary condition is the requirement that the operators of system and environment commute.

For these reasons, we represent the states of the total system by a set $\mc{U}$ which is a \emph{subset} of the Cartesian product $\mc{S}\times \mc{E}$ (which subset will depend on the details of the theory under study). We write $s* e$ to be the combination of $s$ and $e$ --- thus, $s* e$ is equal to $\langle s,e\rangle$ when $\langle s,e\rangle\in \mc{U}$ and is undefined otherwise.

In this abstract formalism, the boundary conditions themselves are best represented as sets of states. Given a subsystem state $s$, the boundary condition $B_s$ of that state is the set of all $e\in \mc{E}$ such that $s* e$ is defined. And conversely, the boundary condition $C_e$ of any \emph{environment} state $e$ is the set of all $s\in \mc{S}$ such that $s* e$ is defined.

What of the symmetries? In keeping with the principle that (mathematically, at least) symmetries are defined in the first place for the total system, we suppose that the \emph{theoretical symmetries} of the theory are represented by a set $\Sigma$ of 1:1 maps from $\mc{U}$ onto itself, closed under inverses and composition. We require that these maps restrict uniquely to well-defined 1:1 maps of \mc{S} and \mc{E} onto themselves: this amounts to the requirement that  for all $s \in \mc{S}, e \in \mc{E}$\rm, $\sigma(s* e)=\sigma|_S(s)* \sigma|_E(e)$ for some maps $\sigma|_S$ and $\sigma|_E$. The theoretical symmetries $\Sigma_S$ of \mc{S} and $\Sigma_E$
of \mc{E} are just the sets of all such $\sigma|_S$ and $\sigma|_E$ respectively.

\section{An example: Coulombic electrostatics}
\label{coulombic_electrostatics}

At this point, it will be helpful to see how these formal ideas play out in a concrete example. Over the course of this paper we will develop several such examples, but our first --- and simplest --- is \emph{Coulombic electrostatics}, the theory of charged particles interacting electrostatically (applicable in the real world in a low-velocity classical regime where radiation and magnetic fields can be neglected).

To be precise, we wish to consider the theory of some set of charged particles moving in a static potential field, together with (optionally) some background conducting surfaces whose evolution in time is fixed \emph{ab initio}. The dynamics are given by the Coulomb force law
\begin{equation} \label{coulomb1}
\ddot{\vctr{x}}_i=-\frac{q_i}{m_i}\nabla \Phi (\vctr{x}_i)
\end{equation}
(where $\vctr{x}_i,q_i$, and $m_i$ are the position, charge and mass of the $i$th particle),
by the Poisson equation
\begin{equation}\label{coulomb2}
\nabla^2 \Phi(\vctr{x})=4\pi\rho(\vctr{x})=4\pi \sum_i q_i\delta(\vctr{x}-\vctr{x}_i)
\end{equation}
and, in the presence of background conductors, by the boundary condition that $\nabla \Phi$ is normal to --- and so $\Phi$ is constant on --- any conducting surface, and by some elastic collision law whereby charged particles bounce elastically off surfaces. (Readers allergic to delta functions should give each particle a small finite radius and adjust ($\ref{coulomb2}$) accordingly.) We assume that not all particles have the same charge-to-mass ratio (the significance of this assumption will become clear in section \ref{newtonian_gravity}).

What are the total-system symmetries of the theory? There is an internal (that is, not spacetime) symmetry
\begin{equation}
\Phi(\vctr{x},t)\longrightarrow \Phi(\vctr{x},t)+f(t);\,\,\,\,\vctr{x}_i\longrightarrow \vctr{x}_i.
\end{equation}
In the absence of background conductors, spatial and temporal translations, velocity boosts, and rotations are also all symmetries. Background conductors, in general, break this symmetry.

There are two natural candidates for the role of isolated subsystem. The first is any collection of charged particles located far from the remaining charged particles and from the conducting surfaces. Even if the system and/or environment have a net charge, the effects of the charge of each on the other will be negligible due to the great distance between them. The second is any collection of charged particles located entirely within some conducting surface. The boundary conditions, in either case, are the same: the subsystem and environment potentials must match up on the spatial boundary of the subsystem.

There are two ways to formalise the subsystem/environment split. First, suppose we are interested only in internal symmetries (in particular, as we will be when considering subsystems that are isolated by being contained within a closed conducting surface). In that case, if (intuitively) our subsystem is localised within region \ss, and contains $p$ particles, then we will specify states of the subsystem by a potential function on $\ss$ (at all times) together with the trajectories of the particles in the specified set $P$ ($|P|=p$), all satisfying the Coulomb law and the Poisson equation; and the state of the environment is given by $N-p$ trajectories in, and a potential function on, $R^4-\ss$ (where $N$ is the total number of particles in the universe). States of the total system are given by superimposing the two; provided the two potential functions agree on the boundary of \ss, the result will be a solution to the equations of motion of the total system.

Things are slightly messier when we wish to consider spacetime symmetries in conjunction with a subsystem/environment split that is itself spatiotemporal in character, since then, in general, the transformation we wish to consider will not leave invariant the `subystem region' $W \subset \field{R}^4$. To deal with this, we will represent the subsystem by  the trajectories of the particles in the set $P$ within, and one potential function defined on, the whole of the spacetime; the environment will be represented similarly.
Still, we take some particular region $W \subset \field{R}^4$ to be part of the characterisation of the subsystem/environment split. For given states of subsystem and environment, the corresponding state of the universe is given by  superimposing the trajectories of the subsystem and environment particles, and setting the potential function equal to the subsystem function within \ss  and to the environment function outside $\ss$. If the particles of the subsystem are contained within $\ss$ and far from its boundary, and the particles of the environment are contained within $\field{R}^4-\ss$ and again are far from its boundary, then the potential functions of both system and environment will be very close to constant on the boundary of $\ss$. As such, the result of combining the subsystem and environment states as above will be a solution of the dynamical equations whenever the potential functions take the same constant value on the boundary of $\ss$.\footnote{There are other symmetries for whose analysis even this more complicated definition would be inadequate. If we were interested in \emph{permutation invariance}, for instance, our practice of taking the subsystem to be specified in part by \emph{some particular collection of particles} would be inadvisable. Plausibly, for any (informally) given subsystem/environment split, there is always a most general way of formalising that split, adequate to dealing with all symmetries at once, and coming at the cost of increased complication.}

\section{A framework for symmetries II --- the relationship between theoretical and empirical symmetries}\label{framework2}

We now return to the general problem of constructing a framework in which to study the notion of `direct empirical significance'---that is, to study the relationship between theoretical and empirical symmetries. Although we have defined (theoretical) symmetries in terms of the total system, we are more directly interested, for the purposes of this task, in the symmetries of the subsystem \mc{S}.  We will find it useful to distinguish two classes of subsystem symmetries. As we observed at the start of this paper, a subsystem symmetry can have direct empirical significance only if,  roughly,  the operation of acting on the subsystem with that subsystem symmetry and `leaving the environment alone' is not a symmetry of the whole system. Accordingly, we first define:

\bd
\item[Interior symmetries.] A subsystem symmetry $\sigma_\mc{S}:\mc{S} \rightarrow \mc{S}$ is \emph{interior} iff $\sigma_\mc{S} * \field{I}$ is a symmetry of the whole system: that is, iff for all subsystem states $s \in \mc{S}$ and all environment states $e \in \mc{E}$ such that $s * \mc{E}$ is defined,
    \ben
    \item the state $\sigma_\mc{S}(s) * e$ is also defined, and
    \item $s * e$ and $\sigma_\mc{S} (s) * e$ represent the same possible world as one another.
    \een
\ed

\noindent We write $I_\mc{S}$ for the group of interior symmetries of a given subsystem $\mc{S}$.

Examples of spacetime symmetries that are interior include the diffeomorphisms used in the Hole Argument\footnote{See \citeN{Norton2008} and references therein.} (treating the `hole' as the subsystem region \ss). Examples of internal symmetries that are interior include gauge transformations which coincide with the identity on a neighbourhood of the boundary of the subsystem region.  (Note that the only sense in which a given \emph{universe} symmetry can be said to be `interior' is subsystem-relative (viz. that its restriction to some specified subsystem is interior in the sense of the above definition); although some universe symmetries --- notably the nontrivial global symmetries --- are not interior relative to any subsystem.)

Since performing an interior symmetry transformation on the subsystem state and leaving the environment state alone results in a redescription of the same possible world, such a subsystem transformation does not lead to a \emph{distinct} situation, hence no (nontrivial) empirical symmetry is associated with such transformations.\footnote{In the terminology of \citeN{Kosso2000}: the performance of an interior symmetry transformation satisfies the Invariance Condition, but not the Transformation Condition.}

We now define a collection of subgroups of the group of subsystem symmetries, each of which includes the interior transformations: the group of those subsystem symmetries that `leave the boundary conditions alone' when acting on some specified subsystem state $s \in \mc{S}$.

\bd
\item[Boundary-preserving symmetries.] A subsystem symmetry $\sigma_\mc{S}:\mc{S} \rightarrow \mc{S}$ \emph{preserves the boundary of $s \in \mc{S}$} iff $s, \sigma(s)$ are elements of the same boundary condition as one another: that is, iff for all $e \in \mc{E}$ and all $s * e \in \mc{U}$, we have $\sigma_\mc{S}(s) * e \in \mc{U}$.
\ed

\noindent In general, we expect---although we do not assume---that a subsystem symmetry will be boundary-preserving \emph{on all states} just in case it is an interior symmetry. (For example, in the case of local internal symmetries in a field theory, the only way for a subsystem symmetry $\sigma_\mc{S}$ to be boundary-preserving on all states is for it to coincide with the identity transformation in a neighbourhood of the subsystem boundary, in which case it is the restriction of the symmetry $\sigma_\mc{S} * \field{I}$ to the subsystem $\mc{S}$, and hence is an interior symmetry.)

The more interesting case is that of subsystem symmetries \emph{that are  boundary-preserving on  some set of subsystem states that arise when the subsystem is isolated from the environment, and  in general not otherwise}. For an example of this phenomenon, consider again the case of  boosting a spatially isolated collection $P$ of particles. Recall (from section \ref{coulombic_electrostatics}) that we take a state of the subsystem to be given by the trajectories of the particles in $P$ and an associated potential function (viz. that component of the total potential function that is due to the particles in $P$), all defined on the whole of spacetime. The motivation for picking out the particular particle collection $P$ as a subsystem is that \emph{we are considering} universe states with the feature that $P$ is spatially isolated (i.e. with the feature that the particles in $P$ are contained within  some region $\ss$ and far from its boundary,  while the particles of the environment are contained within $\field{R}^4-\ss$ and again are far from its boundary). The boundary conditions are then given by matching the subsystem and environment potential functions on the boundary of $\ss$. For those states in which the conditions that motivated picking out $P$ as a subsystem in the first place hold, in the vicinity of the boundary there are no trajectories and potentials are flat, and hence Poincar\'{e} transformations that are `small' enough to preserve the fact that the particles are far from the boundary of $\ss$ will preserve the boundary conditions. But it is a merely contingent fact that this collection of particles is isolated: having thus defined our subsystem $\mc{S}$, there are subsystem states $s \in \mc{S}$ according to which the particles are very close to $\d \ss$, and the boundary conditions of \emph{those} states will not be preserved  even by  `small' Poincare transformations.

Many of the notions we are  using, of course, apply only approximately in practice. For example, no collection of particles is ever \emph{entirely} isolated from its environment: however far we place the environment particles from the subsystem boundary, the environmental potential function on that boundary will not be \emph{exactly} the same as it would have been had the environment particles not existed. The salient aspect of this failure of exact isolation for our project is the following: symmetries will rarely if ever \emph{exactly} preserve the boundary conditions of our subsystem states of interest, and (hence) if the subsystem state $s$ and environment state $e$ jointly solve the equations of motion, in general $\sigma_\mc{S}(s)$ and $e$ strictly speaking will not. However, if the subsystem is approximately isolated from its environment, then $\sigma_\mc{S}$ will \emph{approximately} preserve the boundary of $s$, in the sense that there is another state $s^\prime \in \mc{S}$ `close' to $\sigma_\mc{S}(s)$, whose boundary condition is the same as that of $s$. In such circumstances we will simply write $\sigma_\mc{S}(s) * e \in \mc{U}$, understanding $\sigma_\mc{S}(s) * e$ to denote the result of combining $s^\prime$ and $e$.\footnote{In order to conform to the requirement that subsystem and environment states determine total-system state uniquely, we will have to suppose that some particular choice of which `close' solution to pick has been specified, but the details of how this is done will not be of physical consequence.}

The question now is: when do subsystem symmetries have direct empirical significance? Well: as we noted at the outset, this should occur when there is some transformation of the universe that is not a symmetry of the universe, but whose restriction to the subsystem \emph{is} a symmetry of that subsystem. Accordingly, in the terminology just defined:

\ben

\item Interior symmetries cannot have direct empirical significance, since combining an interior symmetry of the subsystem with the identity on the environment gives a universe symmetry transformation.

\item Non-interior symmetries that are boundary-preserving on some subset $\mc{P} \subseteq \mc{S}$ of states of the subsystem can have direct empirical significance when performed on subsystem states in that subset $\mc{P}$: they are the restrictions of (universe) symmetries to the subsystem $\mc{S}$, and combining them with the identity transformation on the environment gives a transformation that preserves the satisfaction of the dynamical equations (by boundary-preserving-ness), but  need not be a (universe) symmetry transformation (by non-interiority). The empirical symmetry in this case will be \emph{purely relational}, in the sense that the intrinsic properties of both subsystem and environment separately are entirely unaffected, and it is only the relations between the two (for example, the relative position, orientation or velocity) that are changed by the transformation.\footnote{In the terminology of \citeN{Kosso2000}, case (1) corresponds to the case where the Transformation condition cannot be met, case (2) to the case where the Transformation and Invariance conditions are both met.}

\item Non-interior symmetries that are \emph{not} boundary-preserving on the subsystem states of interest can have direct empirical significance, but, in order to realise this significance, the environment state must be altered. Firstly, in order to generate a dynamically possible state of the universe at all, we require some transformation $e \mapsto e^\prime$ of the environment with the feature that $e^\prime \in B_{\sigma(s)}$. There is always, of course, a `cheap' way of finding such an $e^\prime$: simply perform $\sigma_\mc{E}$ on $e$, where $\sigma_\mc{E}$ is the restriction to the environment of the \emph{same} universe symmetry transformation whose restriction to the subsystem we are performing on $s$. This cheap trick, however, fails to generate a distinct physical state of the universe, and hence does not lead to a nontrivial empirical realisation of the subsystem symmetry in question. We can find a dynamically possible state of the universe whose restriction to the subsystem is $\sigma_\mc{S}(s)$ \emph{but that represents a distinct possible world from $s*e$} if we can find an environment state $e^\prime$ that differs physically from $e$ (i.e., in general, whose difference from $e$ could be detected by means of measurements performed on the environment alone), while satisfying the condition $e^\prime \in B_{\sigma(s)}$. In this case, the universe states $s * e, \sigma_\mc{S}(s) * e^\prime$ will be physically distinct, but the difference will not be detectable by means of measurements confined to the subsystem --- that is, we will have a nontrivial empirical symmetry associated with the non-boundary-preserving theoretical subsystem symmetry $\sigma_\mc{S}$.\footnote{There is another cheap trick that avoids the problem mentioned above: first, perform $\sigma_\mc{E}$ on $e$; second, make some small and irrelevant, but intrinsic, further change to the state of the environment. The task of demarcating transformations $e \mapsto e^\prime$ that appear to correspond in some interesting way to the theoretical symmetry of interest, on the one hand --- such as, arguably, the physical implementation of a potential shift by placing a charge on the exterior of Faraday's cage --- and implementations of this modified cheap trick, on the other, lies beyond the scope of this paper. It is the project of stating when a given empirical symmetry with the above structure really deserves to be counted as `associated to' the non-boundary-preserving theoretical subsystem symmetry in question.}
\een

In the case of Coulombic electrostatics, for instance:
\begin{enumerate}
\item There are no interior symmetries (except the trivial case: the identity).
\item In the absence of background conducting surfaces, the spacetime symmetries --- the boosts, rotations and translations --- are all non-interior symmetries  that are boundary-preserving when the system is spatially isolated from the environment. These correspond to Galileo-ship type empirical symmetries.
\item The potential shifts are non-interior symmetries but are not boundary-preserving  on any states. The associated empirical symmetry is that of Faraday's cage; the intrinsic change to the environment state that we implement is the placement of an electric charge on the exterior of the cage.\footnote{\citeN{Healey2009} agrees that the Faraday cage empirical symmetry would correspond to potential shifts \emph{in a world that was exactly described by electrostatics}. Hence, \emph{so far}, our treatment does not disagree with his. The disagreement arises because we hold that the Faraday-cage effect \emph{still} corresponds to the potential shifts once the underlying theory is changed from electrostatics (in which the potential shifts are a global symmetry) to electromagnetism (in which they are local). We discuss this latter case in section \ref{electromagnetism} below.}
\end{enumerate}

Actually, it is potentially a little misleading to speak of \emph{a given non-interior symmetry} having direct empirical significance. The reason is that, for any subsystem symmetry $\sigma$ of $\mc{S}$, any environment transformation $\rho: \mc{E} \rightarrow \mc{E}$ \footnote{Not necessarily an environment symmetry. In the cases we are interested in, $\rho$ is \emph{either} the identity transformation on $\mc{E}$ (hence trivially an environment symmetry), or is a transformation whose effect is to tweak the environment state just enough to render it compatible with the transformed subsystem state (for example, by placing a charge on the outside of the cage).} and any \emph{interior} subsystem symmetry $j$ of $\mc{S}$, `performing $\sigma * \rho$' and `performing $j \cdot \sigma * \rho$' amount physically to the same thing as one another. So there is no empirical significance to the fact that it is $\sigma$ \emph{as opposed to, say, $j \cdot \sigma$} that is performed on the subsystem. This motivates the following definition: we say that two subsystem symmetries $\sigma_1, \sigma_2$ are \emph{equivalent} iff they differ by an interior symmetry (that is, iff there exists $j \in I$ such that $\sigma_1 = j \cdot \sigma_2$). Empirical symmetries correspond $1:1$, not to non-interior (theoretical) symmetries themselves, but to \emph{equivalence classes of non-interior symmetries} under this equivalence relation: that is, to elements of the quotient group $\Sigma_\mc{S} / \mc{I}_\mc{S}$. We will, however,continue to write of individual theoretical symmetries `having direct empirical significance' where confusion will not result.

\section{Newtonian gravity}
\label{newtonian_gravity}

We now begin illustrating these ideas for theories other than Coulombic electrostatics. Our first example is another theory with only global symmetries, and actually it looks pretty similar to electrostatics: it is the theory of some set of massive particles moving in a Newtonian gravitational field, so that the dynamics are given by Newton's law
\begin{equation}
\ddot {\vctr{x}}_i=-\mathbf{\nabla} \Phi(\vctr{x}_i),
\end{equation}
where $\vctr{x}_i$ is the position of the $i$th particle, and by the Poisson equation
\begin{equation}
\nabla^2 \Phi(\vctr{x})=4\pi \rho(\vctr{x})=4 \pi \sum_i m_i \delta(\vctr{x}-\vctr{x}_i).
\end{equation}

\noindent Note that these equations are identical to the equations (\ref{coulomb1})--(\ref{coulomb2}) of electrostatics, \emph{except that} in the gravitational case, the `charge-to-mass ratio' is always unity, hence is the same for all particles. This difference, as is well known, leads to interesting and important effects. The point of the present section is to show how such effects are predicted in our framework; the case also serves as a second example of a global symmetry that is non-boundary-preserving, and hence gives rise to an empirical symmetry that is not purely relational.

As in Coulombic electrostatics, we will take the subsystem to correspond to some fixed spatial region \ss, and we will take a subsystem state to be specified by the trajectories of some (fixed) subset of particles together with a potential function over all space, such that they jointly satisfy the dynamical equations; an environment state is specified by the trajectories of the residual particles, together with another potential over all space. A given subsystem state is  compatible with a given environment state if  (a) the subsystem particle trajectories are confined to the subsystem region \ss; (b) the environment particle trajectories are confined to the complement of \ss; (c) the potential functions agree on the boundary of \ss.\footnote{As in our discussion of Coulombic electrostatics, in practice these conditions will only approximately be met.}

In particular, if the subsystem particles are at all times inside \ss and sufficiently far removed from the environment particles then subsystem and environment will be compatible iff their potential functions have the same behaviour at spatial infinity (since, in that case, their behaviour on the boundary of \ss is given by their behaviour at spatial infinity).

The symmetries of electrostatics---Galilean transforms and time-dependent, spatially independent potential shifts---are also symmetries of Newtonian gravitation. The Galilean transforms lead to exactly the same, purely relational, empirical symmetry in both cases. In principle, the potential shifts lead to non-relational empirical symmetries, but in practice, the absence of conducting surfaces makes this of little interest.\footnote{One counter-example is the empirical symmetry between a system of particles far out in space and the same system in the interior of a hollow sphere of massive particles:  as is well known, the gravitational potential of such a sphere is constant within the sphere. }

There is, however, a third group of symmetries, arising from the above mentioned difference between the equations of electrostatics and those of Newtonian gravity. This corresponds to applying a spatially independent but otherwise arbitrary acceleration $\vctr{k}(t)$ to all the particles, beginning at some time $t_0$. In closed form, this takes a particle at point $\vctr{x}(t)$ at time $t$  to
\begin{equation}
\vctr{x}'(t)=\vctr{x}(t)+\int_{t_0}^t\dr{\tau}_1\int_0^{\tau_1}\dr{\tau_2}\vctr{k}(\tau_2).
\end{equation}
If in addition to this transformation of the particle positions we also replace $\Phi$ with $\Phi'$, where
\begin{equation}\label{gravshift}
\Phi'(\vctr{x}',t)=\Phi(\vctr{x},t)+\vctr{k}(t)\cdot \vctr{x},
\end{equation}
then the resultant system still satisfies the equations of motion; the transformation is a symmetry.

What, if anything, is the empirical significance of this third group of symmetries? Restricted to any subsystem, it is not boundary-preserving  on any state, so it leads to no purely relational empirical symmetries.
 As in the case of potential shifts in electrostatics, however, failure to preserve boundary conditions will not prevent our symmetry from having empirical realisations. We merely have to find distinct states $e, e^\prime$ of the environment, such that $s * e \in \mc{U}$ and such that the boundary conditions of $e$ and $e^\prime$ are related to one another by the subsystem symmetry in question. The transformation $s * e \mapsto \sigma_\mc{S} (s) * e^\prime$ will then be an empirical realisation of our symmetry; if $e, e^\prime$ represent intrinsically distinct physical states of the environment then this transformation will lead to a representation of a distinct possible world, but it follows from the fact that $\sigma_\mc{S}$ is a subsystem symmetry that the difference will not be detectable within the subsystem.

It is easy to find such $e, e^\prime$ in the case in which the subsystem and environment are spatially isolated from one another, i.e. the subsystem and environment particles are both far from the boundary of the `subsystem region' \ss. The boundary conditions are the requirements that the values and derivatives of the subsystem and environment gravitational potentials match on the boundary $\d \ss$ of the subsystem region; thus, since the untransformed and transformed subsystem potentials $\Phi_s, \Phi^\prime_s$ are related as in (\ref{gravshift}), the necessary and sufficient condition for matching is that the untransformed and transformed environment potentials $\Phi_e, \Phi^\prime_e$ are also related by
\be
\Phi^\prime_e(\vctr{x},t) = \Phi_e(\vctr{x},t) + \vctr{k} \cdot \vctr{x}.
\ee
That is, the condition is that the transformed state of the environment differs from the untransformed state by the addition of a uniform gravitational field of strength $\vctr{k}$ on $\d \ss$.

The prediction, in other words, is that in Newtonian gravity, a system floating freely in space behaves identically, with respect to its internal processes, to a system freely falling in a uniform gravitational field. The empirical symmetry associated to the symmetry (\ref{gravshift}) is Einstein's elevator, the thought experiment which led Einstein to the equivalence principle.

It is worth emphasizing that the conditions required for realisation of this empirical symmetry are easy to find in nature; sophisticated technology is not required. Since the gravitational field is smooth, it will count as ``approximately uniform'' in \emph{any} given spacetime region, for the purposes of a sufficiently small subsystem (for example, a human being jumping from a diving board in the Earth's gravitational field). More formally, the condition is realised  when, in some solution to the dynamical equations,  there is some spacetime tube of width $L$ for which
\begin{enumerate}
\item At any time $t$, the potential within the tube has sufficiently small second derivative to be approximated as
\be
\Phi(\vctr{x},t)\simeq \Phi(\vctr{x}_0,t)+\nabla \Phi(\vctr{x}_0,t)\cdot (\vctr{x}-\vctr{x}_0)
\ee
\item The centre of the tube is accelerating with acceleration, at time $t$, equal to $-\nabla \Phi(\vctr{x}_0,t)$.
\end{enumerate}
This in turn will occur whenever the tube is a width-$L$ tube surrounding the trajectory of a test particle moving freely under gravity and the tidal effects in this tube can be neglected (\iec when any significant massive bodies are distance $\gg L$ from the tube). In this circumstance, the boundary conditions of the tube match those of the subsystem up to an overall acceleration, so there will be some symmetry transformation of the subsystem which allows it to be superimposed into the tube.

\section{Local symmetries that are not boundary-preserving: Classical electromagnetism and Faraday's cage}
\label{electromagnetism}

We now wish to consider theories with local symmetries: that is, roughly speaking, symmetries specified by functions on spacetime, rather than by global parameters. (We shall have no need for a more precise definition of ``local''.\footnote{The defender of the conventional wisdom, of course, must believe that some more precise account can be given, since \emph{she} believes that the global/local distinction has some fundamental importance for the notion of direct empirical significance. On \emph{our} view, on the other hand, it would not be surprising if nothing more than a rough-and-ready characterisation of the global/local distinction could be given. As it turns out, though, our analysis will suggest one: cf. footnote \ref{def_local}.}) One of the simplest examples of such a theory is classical electrodynamics.

States in electrodynamics are given by the trajectories $x_i$ of a collection of charged particles (with the $i$th particle having charge $q_i$), and by an electromagnetic four-potential, $A_\mu$, defined on spacetime. The dynamics are given by Maxwell's equations
and by the Lorentz force law.

Recall that this theory is the locus of our original disagreement with Healey: all parties agree that the Faraday Cage phenomenon corresponds to global potential shift in Coulombic electro\emph{statics}, but we and not Healey think this continues to hold once the potential-shift symmetry is `localised'.

Our treatment of this case should, by now, be rather easily predictable.
As before, we first list the theory's symmetries. There are global translations, rotations, and Galilean boosts. In contrast to electrostatics and Newtonian gravity, however, the theory of electrodynamics also has a \emph{local} symmetry: the group of `gauge transformations' of the form
\be
\begin{array}{rcl}
A_\mu \rightarrow A_\mu + \partial_\mu \chi,
\end{array}
\ee
where $\chi$  is an arbitrary smooth real-valued function.

We next consider what might count as an `isolated subsystem' in this theory. The natural answers, as in the case of Coulombic electrostatics, are:

\bi
\item Any collection of charged particles located far from the remaining charged particles and from any conducting surfaces (in which case the subsystem region $\ss$ is taken to be some region such that, in the states we are interested in, our subsystem particles' trajectories are contained well within, and the environment particles' trajectories are well outside, $\ss$).
\item Any collection of charged particles located entirely within some conducting surface (in which case $\ss$ is taken to be the region strictly enclosed by the conducting surface).
\ei

Again as in the case of electrostatics, the details of how we represent subsystem and environment states depend on whether our interest is in spacetime symmetries or only in interior symmetries:
\bi
\item If we wish to discuss spacetime symmetries, we specify subsystem and environment states by giving the relevant particles' trajectories in, and a four-potential function $A_\mu$ on, the whole of the spacetime; for given states (in this sense) of the subsystem and environment, the corresponding state of the universe (if it exists) is obtained by superimposing the trajectories of the subsystem and environment particles, and by grafting the environment potential outside $\ss$ and the subsystem potential inside $\ss$.
\item If our interest is only in internal symmetries (as, in the present section, it is), a simpler definition suffices: again as in section \ref{coulombic_electrostatics}, we define the subsystem four-potential only on $\ss$, and the environment potential only on $M - \ss$.
\ei

This is the first theory we have studied which has nontrivial interior symmetries: if  a smooth function $\chi_S:A\rightarrow \field{R}$ vanishes on a neighborhood of the boundary of \ss, then it defines a gauge transformation which is an interior symmetry of the subsystem: it is the restriction to the subsystem of the universe symmetry transformation $\chi:M \rightarrow \field{R}$ defined by
\be
\chi(x)= \begin{cases} \chi_S(x), & x\in \ss\\ 0, & \mbox{otherwise}. \end{cases}
\ee
%\begin{eqnarray}
% \chi(x)&=&\chi_S(x) \,\,\,(x\in A)\\
%\chi(x)&=&0 \,\,\, (\mbox{otherwise})
%\end{eqnarray}
But not all gauge symmetries of the subsystem are interior symmetries: indeed, if $\chi_S$ does not vanish on the boundary, then defining $\chi$ in this way leads to a discontinuity on the boundary of $\ss$, which violates the requirement that $\chi$ is a smooth function. This is not mere mathematical pedantry: a discontinuity in $\chi$ generates a delta-function potential on the boundary, and so (obviously!) does not leave the environment state unchanged.

For the purposes of identifying and classifying the empirical symmetries in physical situations well-described by this theory, we need to identify which  non-interior (theoretical) subsystem symmetries are and are not  boundary-preserving, relative to subsystem states that satisfy the conditions for the subsystem to be isolated. The answers are that
\bi
\item the Poincar\'{e} transformations, as usual, are boundary-preserving on states satisfying the conditions for spatial isolation, and give rise to Galileo-ship type empirical symmetries;
\item gauge transformations (except those that vanish in a neighbourhood of the subsystem boundary, hence are `interior' and have no empirical significance) are non-boundary-preserving; those whose restriction to the subsystem boundary $\d A$ coincide with (the restrictions of) constant potential shifts correspond to the Faraday cage empirical symmetry (where, as in our discussion of electrostatics, the environment state must be intrinsically changed in order to obtain a nontrivial empirical realisation of the symmetry).
\ei

The shift from the global `gauge' group of electrostatics to the local group of electrodynamics thus has no great effect on the empirical status of the potential shifts: its only effect is that, in the case of the local group, it is strictly speaking the equivalence classes of gauge transformations that tend to the same constant value as one another on the subsystem boundary (elements of $\Sigma_\mc{S} / \mc{I}_\mc{S}$), rather than the individual gauge transformations themselves (elements of $\Sigma_\mc{S}$), that have empirical significance.\footnote{\label{def_local} This point can be ignored when only global symmetries are present, since in these cases there are no nontrivial interior symmetries (i.e. $\mc{I}_\mc{S} = \{ \field{I} \}$), and hence $\Sigma_\mc{S} / \mc{I}_\mc{S}$ is canonically isomorphic to $\Sigma_\mc{S}$. In fact, we can take the existence of nontrivial interior symmetries to \emph{define} what it is for a symmetry to be `local' rather than `global' (relative to a given subsystem/environment split); this may be preferable to the problematic talk of `symmetry groups parametrised by arbitrary functions of spacetime' that is usually used in attempts to define `local'.\label{defining_local}}

Our conclusion is thus (\emph{pace} Healey) that the Faraday cage effect \emph{is} a direct consequence of the local theoretical gauge symmetry of classical electromagnetism in just the same way that other empirical symmetries are direct consequences of global theoretical symmetries, and in just the same way that Faraday's cage \emph{itself} is a consequence of the global gauge symmetry of electrostatics. Our argument is simply that the general framework that we think correctly captures the phenomenon of `direct empirical significance', viz., the framework developed in this paper, straightforwardly entails this result. Since the symmetry in question is non-boundary-preserving, and hence the corresponding empirical symmetry not purely relational, however, the case to which Faraday's cage is directly analogous is that of Einstein's elevator, rather than Galileo's ship.

\section{Local boundary-preserving symmetries: Klein-Gordon-Maxwell gauge theory and t'Hooft's beam-splitter}
\label{kgm}

We turn now to another theory with `local' symmetries: Klein-Gordon-Maxwell theory, the theory of a complex scalar field minimally coupled to a U(1) connection. Here our task is to justify the claim we made in section \ref{faraday_thooft}: that t'Hooft's beam-splitter is a \emph{direct} analog of Galileo's ship, more so even than Faraday's cage.

Mathematically speaking, states of Klein-Gordon-Maxwell theory are specified by a four-potential $A_\mu\equiv(\vctr{A},\Phi)$ and a complex  scalar function $\psi$, and the dynamics are generated by the Lagrangian density
\be
\mc{L}= \left( (\partial_\mu-iqA_\mu)\psi \right)^* \left(\partial_\mu-iqA_\mu \right)\psi +m^2 \psi^*\psi+\mc{L}_{EM}, \label{kgm_lagrangian}
\ee
where $\mc{L}_{EM}$ is the standard Lagrangian of classical electrodynamics in the absence of matter. This theory can be understood either as a classical field theory (albeit one not realised in nature) or, as is more usual, the semiclassical theory of a quantized charged scalar particle (a pion, for instance) interacting with a classical background electromagnetic field.

The symmetries of this theory include Poincar\'{e} boosts, translations and rotations, which lead to Galileo's ship-type empirical consequences as usual. More interestingly for our purposes, it also has the `gauge' symmetry
\bea
\psi(\vctr{x},t) & \mapsto & \exp(-i q \chi(\vctr{x},t))\psi(\vctr{x},t) \\
A_\mu(t,\vctr{x}) & \mapsto & A_\mu(t,\vctr{x}) + \d_\mu \chi(t,\vctr{x}),
\eea
where $\chi$ is an arbitrary smooth function. It is easily verified that such joint transformations leave invariant the Lagrangian (\ref{kgm_lagrangian}), and hence all equations of motion derived therefrom.

The natural choice of isolated subsystem is a system of matter and electromagnetic fields far from other matter concentrations and from radiation.
We specify a subsystem by giving a set $\ss \subset \field{R}^4$, and the states of subsystem and environment respectively by giving scalar fields and connection on $\ss$ and $\field{R}^4-\ss$, satisfying the dynamical equations in each case. A subsystem state is compatible with an environment state  iff the fields of the two agree on some neighbourhood of the boundary of \ss.

As in the case of electrodynamics, a subsystem symmetry transformation, given by a smooth real-valued function $\chi_S$ on the subsystem region $\ss$, need not be interior; it will be interior only if $\chi_S$ vanishes in a neighbourhood of the boundary $\d \ss$ of the subsystem region. In particular, suppose that $\chi_S$ is constant, but non-zero, on a neighborhood of the boundary. Then (given that we are considering states in which the subsystem is isolated, and so (in particular)  $\psi$ is zero on the boundary) the transformation defined by $\chi_S$ leaves the boundary conditions of our states of interest invariant: that is, it defines a  symmetry that is boundary-preserving on those states. But $\chi_S$ is not an interior symmetry: it does not leave invariant the boundary conditions of \emph{other} subsystem states, in which $\psi$ is \emph{non}zero on the boundary of \ss.  Any two  subsystem symmetries are (by definition) equivalent when they differ by an interior symmetry: so $\chi_S$ and $\chi'_S$ define equivalent symmetries iff $\chi_S-\chi'_S$ vanishes on a neighborhood of the boundary. Assuming that the boundary is connected, this entails that an equivalence class of these symmetries is specified by a real number: the constant value of $\chi_S$ on the boundary. A natural choice of representatives for these equivalence classes are the constant functions, sometimes called the \emph{global} gauge transformations.

According to the general theory of empirical symmetries developed in sections  \ref{framework1} and \ref{framework2}, this means that `local' subsystem gauge transformations that are constant in a neighbourhood of the subsystem boundary correspond,  for subsystem states in Klein-Gordon-Maxwell theory satisfying the conditions for spatial isolation, to purely relational realizations of the gauge symmetry; and we might ask what, if any, are the empirical consequences of these realizations. But the answer to this question is clear:  at least mathematically speaking, the theory of one particle in an electromagnetic field,  described in section \ref{thooft} and there asserted adequate to model t'Hooft's modified beam-splitter, just is  the Klein-Gordon-Maxwell theory discussed in the present section.
Hence the empirical symmetry obtaining between the unshifted and shifted half-beams in t'Hooft's beam splitter is a direct empirical consequence of local gauge symmetry, in the same way that Galileo's ship is a direct empirical consequence of boost invariance.

Our analysis is in conflict with \citeN{bradingbrownsymmetry}: as we have noted, they  argue that interferometry experiments of this kind are not empirical consequences of either global or local symmetries. They reason as follows:
\begin{quote}
[We could] consider a region where the wavefunction can be decomposed into two spatially separated components, and then \ldots apply a local gauge transformation to one region (\ie to the component of the wavefunction in that region, along with the electromagnetic potential in that region) and not to the other. But then either the transformation of the electromagnetic potential results in the potential being discontinuous at the boundary between the `two subsystems', in which case the relative phase relations of the two components are undefined (it is meaningless to ask what the relative phase relations are), or the electromagnetic potential remains continuous, in which case what we have is a special case of a local gauge transformation on the entire system --- and this of course brings us back to where we started --- such a transformation has no observable consequences. (\citeN[p.\,656]{bradingbrownsymmetry})
\end{quote}

Brading and Brown's argument is essentially a more careful version of the argument-sketch we gave in the introduction for why local symmetries might be thought to have no empirical consequences; and, being more careful, it allows us to see why that argument fails. It assumes, tacitly, that if some smooth transformation $\varphi$ of the fields on a (connected) region $A\cup B$ restricts to gauge transformations $\varphi_A$ and $\varphi_B$ on the fields on $A$ and $B$ respectively, then the overall transformation must be a gauge transformation. This would indeed be the case if the physically given transformation $\varphi$ were itself a \emph{function from spacetime to the gauge group}, with $\varphi_A = \varphi \vert_A$ and $\varphi_B = \varphi \vert_B$: for, in that case,
\be
\varphi(x) = \begin{cases} \varphi_A(x), & x \in A, \\ \varphi_B(x), & x \in B, \end{cases}
\ee
 would just be the gauge transformation in question. The key to seeing why this argument fails is noting that what is given, when we are given the pre- and post-transformed states of the universe, is \emph{not} a function from spacetime to the gauge group, but merely the effect of whatever transformation is being performed on the \emph{particular} pre-transformation  (universe) state $(\psi, A_\mu) \in \mc{U}$. And if this particular $\psi$ happens to vanish on the overlap region $A \cap B$, then nothing about the corresponding gauge transformations $\varphi_A, \varphi_B$ can be `read off' from their effects on the wavefunction in that region. It is therefore possible that the universe transformation being performed might correspond to the effect of (say) some constant gauge transformation $\varphi_A$ in \ss, and a \emph{different} constant gauge transformation $\varphi_B$ in $B$, so that there is no way of patching $\varphi_A$ and $\varphi_B$ together to obtain a single smooth function from the whole of spacetime to the gauge group. This possibility is exploited by the scenario under discussion.\footnote{Brading and Brown raise further problems for our analysis: they claim that it is not legitimate to view one half of the split beam as a \emph{subsystem}, which of course is precisely what we are doing. Explaining why they think this is illegitimate, Brading and Brown (writing $\Psi, \Psi_I, \Psi_{II}$ for (respectively) the overall system wavefunction and the components of that wavefunction with support in the region of the upper and lower half-beams, so that $\Psi = \Psi_I + \Psi_{II}$) write:
\begin{quote}
The crucial issue here is whether the two components of the wavefunction, $\Psi_I$ and $\Psi_{II}$, can be interpreted as representing genuine subsystems of $\Psi$. Our position is that only $\Psi$ represents a physical system, with $\Psi_I$ representing one (basis-dependent) component of the wavefunction $\Psi$. \ldots The point is made particularly vivid by considering a single electron passing through the two slits: on our view, there is only one system here, described by $\Psi$, and the components $\Psi_I$ and $\Psi_{II}$ do not represent subsystems of the electron. The same general point holds even on an ensemble interpretation, and --- at least in the absence of further argument --- it seems that $\Psi_I$ and $\Psi_{II}$ cannot be interpreted as representing genuine subsystems.
\end{quote} We find this unpersuasive. For one thing, though the theory is quantum-mechanical, it is mathematically identical to a classical theory in which systems can be understood in precisely the way required; for another, the whole problem can be modelled perfectly effectively using Fock-space methods if we wish to understand the two halves of the beam as two quantum-mechanical systems in the usual sense. But more importantly, `subsystem' is just a word, to be used in whatever way is most perspicuous, and in \emph{this} case (as is not the case in, say, analysis of entanglement), the most perspicuous definition of `subsystem' does indeed seem to treat the two components as distinct subsystems.}

\section*{General relativity}
\label{gr}

Our final example is another ``local'' theory: the General Theory of Relativity, with or without matter fields present. We formulate the theory in the traditional way, in terms of a metric on a manifold; the theoretical symmetries of the theory include all the diffeomorphisms of the manifold. (We ignore any internal symmetries of the matter fields.)

As usual, we have some freedom over what to pick out as a `subsystem'. For our current purposes, we can proceed in much the same way that we did for the analysis of spacetime symmetries in sections \ref{coulombic_electrostatics}--\ref{kgm}.
Thus: subsystem and environment are each specified by giving the fields on all of some (common) underlying manifold $M$, but the subsystem is defined by some particular region $\ss \subset M$.\footnote{Arguably, this definition fails to take into account certain particular features of general relativity: notably (a) the possibility of topology change, so that a subsystem may not have a fixed topology invariant across all states of that system, and (b) the fact that a given subset of the bare manifold does not correspond to anything physical in the absence of the (matter and metric) fields. We ignore these issues for simplicty (see, however, \citeN{wallacerpep}.)}
 A given subsystem state $s$ and environment state $e$ are \emph{dynamically compatible} iff the metric and matter fields of $s$ and $e$ coincide in some neighbourhood of $\d W$; subsystem and environment are mutually \emph{isolated} if (further) the matter fields vanish, and the metric is flat, on such a neighbourhood. We obtain `subsystem symmetries' by allowing diffeomorphisms (of the entire manifold $M$) to act only on the subsystem state. Applying our usual classification of such subsystem symmetries, we see that:
\bi
\item The \emph{interior} symmetries of $\mc{S}$ are those diffeomorphisms that coincide with the identity diffeomorphism on a neighbourhood of $\d \ss$.
\item Diffeomorphisms that coincide with a Poincare transformation on a neighbourhood of $\d \ss$ are non-interior, but are  boundary-preserving on states that meet the condition for isolation. (As usual for spacetime symmetries, we also need to require that these transformations are not large enough to move the non-flat part of the region $\ss$ across the boundary.) These correspond empirically to boosts, rotations and translations. (Note that these operations are only definable in GR where the system being transformed lies in flat spacetime --- \iec, when it is an isolated subsystem in our sense.) So the Galileo-ship empirical symmetry can be seen as a direct empirical consequence of the diffeomorphism symmetry of the theory.
\item \emph{Arbitrary} diffeomorphisms are in general non-interior, and do not preserve the boundary of any states. However, recall, from our general framework, that this need not prevent them from having direct empirical significance: given \emph{any} two subsystem states $s, \sigma_{\mc{S}}(s)$ that are related by a subsystem symmetry, we can find an empirical realisation of this symmetry by finding intrinsically different environment states $e, e^\prime$, such that $s * e \in \mc{U}$ and the boundary conditions of $e, e^\prime$ are related to one another by the same symmetry (so that $\sigma_\mc{S}(s) * e^\prime \in \mc{U}$ also).
\ei

In fact, Einstein's elevator can be seen as an empirical realistation of certain of the non-interior diffeomorphism symmetries of the theory. For 
in general relativity, if 
$u_1$ and $u_2$ are solutions of the equations such that for some  region $\ss \subset M$:
        \ben
        \item $\gamma_1, \gamma_2$ in $\ss$ are geodesics according to $u_1,u_2$ respectively;
        \item spacetime is flat on lengthscales $\sim L$ around $\gamma$ in $u_1$ and around $\gamma_2$ in $u_2$;
        \item tubes $S_1$ around $\gamma_1$ and $S_2$ around $\gamma_2$, each of width $L$, are both contained in $\ss$,
        \een
    then there will be some diffeomorphism $d$, taking the surface of $S_1$ to that of $S_2$, such that 
$d_*u_1|_{\d W} = u_2 |_{\d W}$.
 Thus $u_1,u_2$ restrict to environment states $e_1,e_2$ respectively whose boundary conditions are indeed related by a diffeomorphism, and hence for any subsystem state $s$ compatible with $e_1$, there is a subsystem-symmetry-transformed counterpart 
$d_*s_1$ 
that is compatible with $e_2$.

Which subsystem states are compatible with environment states that satisfy (1)--(3)?  The answer is essentially the same as in Newtonian gravity: any isolated system of matter and metric --- some collection of material bodies, say, or a collection of black holes interacting in some complex manner,  --- of effective size $L$, so that outside a tube of spatial diameter $\sim L$ the metric is approximately flat and the matter fields approximately vanish. (If the isolated system is not strongly self-gravitating, the second condition will entail the first.) Any such system is compatible with  $e_1$, and hence with $e_2$. Hence, any such system will behave identically in any spacetime region whose curvature can be neglected on scales of $\sim L$. And this, of course, is just the (strong) equivalence principle.\footnote{Note that this analysis of the equivalence principle does not in any way require that the spacetime \emph{interior} to the subsystem be approximately flat: the equivalence principle, like the relativity principle, applies to strongly self-gravitating systems, not just systems in which gravity can be neglected (something which we also saw to be the case in Newtonian graviation. See \citeN{wallacerpep} for further discussion.}

\section{Summary}
\label{summary}

We have developed our framework a little at a time, in order to display the motivations for and applications of each of its elements. Here we give a summary of the finished framework itself, for convenience of reference.

\bi
\item Let $\mc{U}$ be the set of dynamically allowed states for the universe (where `state' is intended in the mathematical, rather than the physical, sense).

\item Let $\Sigma$ be the set of symmetries $\sigma: \mc{U} \rightarrow \mc{U}$ of the  universe \rm.

\item Let $\mc{S}$ and $\mc{E}$ be the sets of dynamically allowed states of (respectively) subsystem and environment, and $\pi_\mc{S}:\mc{U} \rightarrow \mc{S}, \pi_\mc{E}:\mc{U} \rightarrow \mc{E}$ the projections from universe states onto (respectively) system and environment states.

\item Require that the structure $(\mc{U}, \mc{S}, \mc{E}, \pi_\mc{S}, \pi_\mc{E})$ satisfies the following conditions:
	\bd
	\item[A1.] For each $s \in \mc{S}, e \in \mc{E}$, there is at most one $u \in \mc{U}$ such that $\pi_\mc{S}(u)=s$ and $\pi_\mc{E}(u)=e$. (The states of subsystem and environment jointly determine the state of the total system.)
	\item[A2.] For all $\sigma \in \Sigma$ and all $u, u^\prime \in \mc{U}$, if $\pi_\mc{S}(u)=\pi_\mc{S}(u^\prime)$ then $\pi_\mc{S}(\sigma(u))=\pi_\mc{S}(\sigma(u^\prime))$; similarly for $\pi_\mc{E}$ . (The projections from the total system to the subsystem/environment respect symmetries.)

	\ed

\item With these requirements in place, we make the following further definitions and observations:
	\bi
\item Define the compatibility relation between $\mc{S}$ and $\mc{E}$ as follows: $s$ is compatible with $e$  iff for some $u\in\mc{U}$, $\pi_\mc{S}(u)=s$ and $\pi_\mc{E}(u)=e$. (This is intended to be read as `the subsystem state $s$ and the environment state $e$ are dynamically compatible' --- \iec (for field equations) their fields approximately match on the boundary between them.)

	\item Define a partial function $*$ from $\mc{S} \times \mc{E}$ onto $\mc{U}$ as follows: if there exists $u \in \mc{U}$ such that $\pi_\mc{S}(u)=s$ and $\pi_\mc{E}(u)=e$, then $s * e = u$; otherwise $s * e$ is not defined. (A1 ensures that this does indeed well-define $s * e$ on the desired domain.)

	\item For arbitrary $s \in \mc{S}$ (resp. $e \in \mc{E}$), define the \emph{boundary condition of $s$} (respectively, of $e$), $B_s$ (resp. $C_e$), to be the set of all $e \in \mc{E}$ (resp. $s \in \mc{S}$) such that $s * e$ is defined.

	\item Define the restriction $\sigma_\mc{S}$ (resp. $\sigma_\mc{E}$ of a given total-system symmetry $\sigma$ to the subsystem (resp. to the environment) as follows: for all $s \in \mc{S}$, $\sigma_\mc{S}(s) = \pi_\mc{S}(\sigma(u))$, where $u \in \mc{U}$ is such that $\pi_\mc{S}(u) = s$. (A2 ensures that $\sigma_\mc{S}(s)$, thus defined, is independent of the choice of representative $u$.) Write $\Sigma_\mc{S}$ (resp. $\Sigma_\mc{E}$) for the set of all subsystem (resp. environment) symmetries.
	\ei

\item We are now in a position to define the notions of an `interior' and a `boundary-preserving' subsystem symmetry:

	\bi

	\item Let $\mc{I}_\mc{S} \subset \Sigma_\mc{S}$ be the set of `interior' symmetries of $\mc{S}$; that is, the set of subsystem symmetries $\sigma_\mc{S} \in \Sigma_\mc{S}$ such that $\sigma_\mc{S} * \field{I} \in \Sigma$.

	\item For any $s \in \mc{S}$, let $\mc{BP}_\mc{S}(s)$ be the set of subsystem symmetries that `preserve the boundary of $s$', i.e. the set of subsystem symmetries $\sigma_\mc{S}$ such that $s, \sigma_\mc{S}(s)$ lie in the same boundary condition $C_e$ as one another. (Clearly, for any $s \in \mc{S}$ we have $\mc{I}_\mc{S} \subset \mc{BP}_\mc{S}(s)$; equally clearly, the interior symmetries are just those symmetries that are boundary-preserving on \emph{all} states $s \in \mc{S}$.)
	\ei
\ei

We take \emph{any} $\left( \mc{U}, \mc{S}, \mc{E}, \pi_\mc{S}, \pi_\mc{E} \right)$ satisfying (A1) and (A2) to define a legitimate decomposition of the universe into `subsystem' and `environment'. Without further restricting the nature of the `subsystem', we wish to indulge in talk of the `intrinsic properties of the subsystem', and so forth; such talk is to be cashed out as follows. (1)
%we take it that
The `intrinsic properties of the subsystem'  must depend on $\pi_\mc{S}(u)$ alone (and not on any other features of $u$). (2) If some transformation $u \mapsto u^\prime$ does not change which possible world is being represented (i.e., if $u$ and $u^\prime$ are related to one another by a universe symmetry), then the intrinsic properties of the subsystem (resp. environment) according to $u$ must be the same as those according to $u^\prime$. (3) an `experiment confined to the subsystem' is an experiment whose outcomes (or probabilities for outcomes) depend on the intrinsic properties of the subsystem alone.

We are now in a position to talk about the conditions under which the intrinsic properties of the subsystem might have \emph{changed}, without having said anything (else) about what those intrinsic properties \emph{are}. Specifically,  the following are now direct consequences of our definitions:
\bi
\item Interior symmetries have no direct empirical significance: if $\sigma_\mc{S} \in \mc{I}(\mc{S})$, then for all $s * e \in \mc{U}$, $\sigma_\mc{S}(s) * e$ is defined \emph{but}
represents the same possible world as $s * e$.
\item If $\sigma_\mc{S}$ is a non-interior symmetry that is boundary-preserving on some state $s \in \mc{S}$, then, for all $e \in \mc{E}$ such that $s * e$ is defined, the states $s * e$, $ \sigma_\mc{S}(s) * e$  in general represent different possible worlds from one another \emph{but}, given that the two subsystem states are related by a subsystem symmetry, then (by (1) and (2)) the transformation induces no \emph{intrinsic} change to the subsystem (and of course induces none to the environment). Hence, the difference will not be detectable by experiments confined to either subsystem or environment alone: all that has been changed are \emph{relations} between the subsystem and environment.
\item Quite generally, if $\sigma_\mc{S}$ is a non-interior symmetry of the subsystem $\mc{S}$ (not necessarily boundary-preserving on any salient collection of subsystem states), still any two environment states $e, e^\prime$ whose boundary conditions are related to one another by $\sigma_s$  are dynamically compatible with the same \emph{intrinsic} states of the subsystem as one another. Formally:  $\forall \sigma_\mc{S} \in \Sigma_\mc{S}, e, e^\prime \in \mc{E}$, $\forall s \in \mc{S}$, if $C_{e^\prime}=\sigma_\mc{S}(C_e)$ then $\forall s \in \mc{S}, s * e \in \mc{U} \Rightarrow \sigma_s(s) * e^\prime \in \mc{U}$. If $e, e^\prime$ are \emph{intrinsically} different states of the environment, then the transformation $s*e \mapsto \sigma_\mc{S} * e^\prime$ is a change in physical state of the universe, and one that is  \emph{not} purely relational.  The empirical symmetry will nevertheless `correspond to' the theoretical subsystem symmetry $\sigma_\mc{S}$, insofar as (and \emph{only} insofar as) there is a principled connection between the subsystem symmetry transformation $\sigma_\mc{S}$ and the environment transformation $e \mapsto e^\prime$.  We have not investigated the possibilities for such `principled connections'. \rm
\ei

\section{Conclusions}
\label{conclusion}

We began this paper with two pieces of conventional wisdom: that the direct empirical consequences of symmetry transformations arise when only a subsystem is transformed, and that because of this, global but not local symmetries have direct empirical significance. In developing an appropriate framework to explicate the first of these, we have found fault with the latter. The perspicuous distinction for analysis of phenomena on the pattern of Galileo's ship is subsystem-relative, and is between symmetries  of the subsystem in question that are \emph{interior} and those that are non-interior, rather than between  restrictions of global and  restrictions of local symmetries  to the subsystem: it is the interior symmetries of a given subsystem that cannot have empirical consequences when performed on that subsystem. The second piece of conventional wisdom is simply false: an element of a given `local' symmetry group would be unable to have empirical significance \emph{simpliciter} only if its restrictions to \emph{all} subsystems were interior symmetries of those subsystems, but no transformation (other than the identity, of course) has this property. The relationship between the interior/exterior and the global/local distinction is this: only when the group of theoretical symmetries is `local' will there \emph{be} any interior symmetries, distinct from the identity, for subsystems that are identifiable as subregions of spacetime.

We are now in a position to resolve the issues raised by the three objections to which the claim that `global symmetries have empirical significance but local symmetries do not' gave rise, at the start of this paper.

Our first objection was that, since the global symmetries form a subgroup of the local symmetry group, it is logically impossible that all global and no local symmetries can have empirical significance. The short way with this objection, in the light of our analysis, is to note that we have rejected the offending claim that local symmetries cannot have empirical significance. However, we should also check that no analogous paradox recurs for our new set of claims. Indeed it does not. The reason lies in a structural difference between the original problematic claims and our replacements: rather than holding that all elements of a (`local') symmetry group have one property while elements of some subgroup thereof have a contradictory property, we hold that (for any given subsystem) there is a subgroup of subsystem symmetry transformations (the `interior' ones) that cannot have empirical significance, and that it is elements of the \emph{quotient} of the larger group by this subgroup that are candidates for correspondence to physical operations. There is thus no object of which we assert both that it does, and that it does not, have some given property.

We will return to our second objection in a moment. Our third objection was more concrete: it was that several particular cases, for example those of Faraday's cage and t'Hooft's beam-splitter, seemed (at least) to be perfect analogs of the Galileo-ship thought-experiment for various local symmetries: that is, cases in which \emph{local} symmetries corresponded to empirical effects in just the same way that the boost invariance of Newtonian mechanics corresponds to the Galileo-ship effect. Our framework classifies the various cases we have discussed as follows:

\noindent
\begin{tabular}{|p{1.3in}|p{0.9in}|p{0.5in}|p{0.4in}|p{1in}|}
\hline
Theory & Theoretical symmetries & Symmetry group global or local? & Boundary-preserving? & Corresponding empirical symmetries \\ \hline \hline
All flat-spacetime theories (e.g. Newtonian mechanics, Coulombic electrostatics, Newtonian gravity) & Galilean/Poincare transformations & Global & Yes & Galileo's ship \\
\hline
General relativity & Asymptotically Poincare transformations & Local & Yes & Galileo's ship \\ \hline
Quantum mechanics with only global phase symmetry & Constant phase shifts & Global & Yes & t'Hooft's beam-splitter \\ \hline
Quantized Klein-Gordon-Maxwell theory & Asymptotically constant gauge transformations & Local & Yes & t'Hooft's beam-splitter \\ \hline
Theories with a `global' potential shift symmetry (e.g. Coulombic electrostatics, Newtonian gravity) & Spatially constant potential shifts & Global & No & Faraday's cage (or gravitational analog) \\ \hline
Theories with a `local' potential shift symmetry (e.g. Classical electrodynamics, general relativity) & Asymptotically constant potential shifts & Local & No & Faraday's cage (or gravitational analog) \\ \hline
Newtonian gravity & Spatially constant accelerations & Global & No & Einstein's Elevator \\ \hline
General relativity &  Diffeomorphisms not vanishing on the boundary of a system & Local &  No & Einstein's elevator\\ \hline
\end{tabular}

\vspace{12pt}
Examining this list, we see that  there is no particularly important distinction between theories with a global symmetry group and those with a local symmetry group, from the point of view of the connection between theoretical and empirical symmetries. Every `global' theory and the corresponding `local' theory will entail exactly the same empirical symmetries, with constant transformations in the former case replaced by equivalence classes of transformations in the latter. (Healey might just as well have posed his puzzle in terms of the ability of general relativity to explain Galileo's ship: the association of Galileo's ship more closely with global symmetries, and Faraday's cage more closely with local symmetries, is spurious.) This in turn resolves our \emph{second} original puzzle, about whether explanatory power is lost when we `localise' a symmetry, in a way that (in particular) should worry scientific realists: it is not.

There \emph{is}, however, a principled distinction between the (non-interior) symmetries that are `boundary-preserving' on states satisfying the condition for dynamical isolation, and the (non-interior) symmetries that are \emph{not} boundary-preserving even on such special states. For this reason, it is t'Hooft's beam-splitter that is \emph{precisely} analogous to Galileo's ship, while Faraday's cage is precisely analogous to empirical realisations of the Equivalence Principle.

\section*{Acknowledgements}

Thanks are due to audiences at various venues at which embryonic versions of this paper were presented: Oxford, LSE, York, Dubrovnik and Cambridge. We are particularly grateful to Katherine Brading and Richard Healey, for illuminating and very constructive discussions.

\end{document}